# Stability of Local Information based Centrality Measurements under Degree Preserving Randomizations

*Chandni Saxena, M.N.Doja, *Tanvir Ahmad

Department of Computer Engineering,
Jamia Millia Islamia,
New Delhi, India
{*cmooncs@gmail.com,ndoja@yahoo.com,*tahmad2@jmi.ac.in}

**Abstract.** Node centrality is one of the integral measures in network analysis with wide range of applications from socio-economic to personalized recommendation. We argue that an effective centrality measure should undertake stability even under information loss or noise introduced in the network. With six local information based centrality metric, we investigate the effect of varying assortativity while keeping degree distribution unchanged, using networks with scale free and exponential degree distribution. This model provides a novel scope to analyze stability of centrality metric which can further finds many applications in social science, biology, information science, community detection and so on.

## 1 Introduction

Relationship among nodes in the network has varied meaning and analogy. These relations are observed in different domains and fields, such as; relationship among neurons in neural network, among web-pages in web graph, among online users in social relationship graph and among authors in scientific collaborations network. Complex networks have been explored to study these relations at microscopic level of elements like nodes and their relations as links. Structural centralities are such measures which uncover the explicit roles played by the nodes and the links. The concept of centrality is characterization of a node, subject to its importance according to the structural information of the network. Centrality has become an important measurement with vast implications in theoretical research such as physical science [1] or biological science [2], and practical significance in applications such as e-commerce [3] and social networks [4]. However, the node centrality concept is vast and can be furnished in various ways, putting forward multiple coexisting centrality measures. Among various aspects, centrality based on local information is of greater importance due to its low computation complexity and simplicity. Also in the evident scenario of real complex networks having incomplete information or loss of information and dynamically changing topological behavior of the network, local information based centralities provide best solutions. The important issue here is to have ability of this metric to be stable and robust to noise in the network or to any randomization, such created under real situations. It has been proven that different network topologies such as scale-free, exponential and heavy-tailed degree distribution [5] can affect network based operations such as robustness of network [6] and epidemic spreading [7]. So far the issue of topological structure has not been investigated for different centralities aiming its stability issue. This has motivated us to explore the stability of local information based centrality with different topologies of the networks. We studied these centralities under scale–free and exponential distribution of networks and examined the effect of network randomization (perturbation) when degree of the nodes was kept preserved. To examine the stability problem in local information based centrality we have analyzed six centrality measurements namely, *H-index* [8] (*h*), *leverage centrality* [9] (*lc*), *local structural entropy* [10] (*lse*), *local clustering coefficient* [11] (*lcc*), *topological coefficient* [12] (*tc*) and *local average connectivity* [13] (*lac*).We provide a framework to categorize these metrics in a novel dimension. In our experiments we vary assortativity of the network leaving the crucial property of degree distribution unchanged according to **noise model 1**. We evaluate these metrics on the basis of average rank difference and standard deviation of average rank difference for original and randomized networks. We also compare overlaps of sets of top ranked nodes, varying range of perturbation with the original network as explained in **noise model 2** defined by Yang [21]. We evaluate this model empirically using real and synthetic networks under scale free and exponential degree distribution.

The rest of the paper is organized as follows. Section 2 presents framework of the stability issues covering theoretical details and related works in this direction. In section 3 results and experiments evaluate the different stability performance metric on different datasets, and finally paper is concluded in section 4.

## 2    Background

### 2.1    Local Information Based Centrality Measure

There have been dozens of centralities based on global and local topological information of a network. A local network around each node in terms of its connectedness, topology and type form a network which can formulate its influence. We have examined six local information based measures which require only a few pieces of information around the node, defined in this section. *H-index* of a node $v_i$ is the largest value h, such that $v_i$ has at least h neighbors of degree no less than h. *Leverage centrality* finds nodes in the network which are connected to more nodes than their neighbors and determines central nodes. *Local structural entropy* of a node is based on Shannon entropy of local structure which depends on degree of its neighboring nodes in the local network of the target node. *Local clustering coefficient* of a node is the ratio of number of links among its neighbors to possible numbers of links that could formulate. *Topological coefficient* of a node is the sum ratio of number of neighbors common to a pair of nodes among directly linked neighbors to the number neighboring nodes of target node. *Local average connectivity* of a node is the sum of local average connectivity of its neighbors in the sub network induced by the target node.

## 2.2 Degree Preserving Randomization

Rewiring network randomly while keeping its degree distribution hence degree of each node constant is to enact real time perturbation of the network that mapped incompleteness or state of loss of information. As local information based centrality measurements are degree dependent, therefore preserving degree of each node while perturbation is an essential step. Primarily this leads to change in mixing pattern of a network and hence changing the assortativity of a node. A pair of edges are picked randomly and rewired as illustrated in figure 1, ensuring degree preserving randomization e.g. edges AC and BD are picked as candidates, now the connections between end points are exchanged to form new edges AD and BC. To study the effect of randomization on centrality stability of the network nodes, we use **noise model 1** and **noise model 2** on six different data sets in this paper. These datasets differ both in size and subject to type as the scale free and the random exponential networks as shown in table I. The *facebook* dataset consist of nodes as people with edges as friendship ties extracted from social network facebook. The *brain-network* is medulla neuron network of fly-drosophilla. The *foodweb* is the food web network data of Little Rock Lake, Wisconsin USA. The *c. elegans-metabolic* is a metabolic network data. Synthetic scale free and synthetic exponential distribution datasets are also considered for the validation of real data. The basic topological features of the networks are listed in table I.

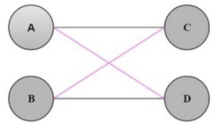

**Fig.1** Network perturbation preserving degree of nodes

## 2.3 Related Work

There are a few numbers of studies on finding stability of pagerank algorithm under randomization [14][15][16][17]. Goshal and Barabasi [18] have proposed stability criteria for scale free and exponential network when degree preserving random perturbation and incompleteness in network are observed. Authors have studied the ranking stability of top nodes and manifested the initial success of pagerank algorithm to scale free nature of the underlying WWW topology. Lempel and Moran [14] have examined the stability of three ranking algorithms - pagerank, HITS and SALSA - to perturbation, while Senanayake et al. [16] have studied the page rank algorithm performance with underlying network topology. Andrew et al. [17] have analyzed the stability of HITS and PageRank to small perturbations. Sarkar et al. [19] investigated community scoring and centrality metric in terms of different noise levels. So far no work has been observed carrying stability study of local information based centralities.

**Table 1.** Topological Statistics of datasets: number of nodes (n), number of edges (m), maximum degree ($k_{max}$), best estimate of the degree exponent of the degree distribution (γ), assortativity of unperturbed network (r) and assortativity of perturbed network ($r_p$).

| Network | n | m | $k_{max}$ | γ | (r) | ($r_p$) |
|---|---|---|---|---|---|---|
| **Scale-free** | | | | | | |
| Synthetic | 2000 | 5742 | 1378 | 2.1 | -0.4157 | -0.4179 |
| Facebook | 6621 | 249959 | 840 | 2.3 | 0.1225 | 0.1149 |
| Brain | 1781 | 33641 | 16224 | 1.9 | -0.3235 | -0.3252 |
| **Exponential** | | | | | | |
| Synthetic | 1000 | 9945 | 495 | - | -0.1252 | -0.1174 |
| C. elegans | 453 | 4596 | 644 | - | -0.0625 | -0.2276 |
| FoodWeb | 183 | 2494 | 108 | - | -0.2374 | -0.1585 |

### 2.4 Stability

To examine the stability of centrality measurements we furnish the data with degree preserving randomization according to two noise models as explained. **Noise Model 1**: We implement number of perturbations in the network of size more than 2000 nodes equal to the size of network and for other networks it is 10 times the size of the network. The reason lies on the fact to ensure the range of perturbation depending upon the size of network and also to realize meaningful change in assortativity of the network. Where **assortativity** is defined as a measure that evaluate the tendency of a node to be connected to similar nodes in the network. Figure 2 reports the distribution of ranking scores for perturbed and unperturbed networks for three centrality measures- *h-index*, *leverage centrality* and *local structure entropy*- on synthetic scale free and exponential network data, real scale free facebook and exponential c. elegans network data. The *h-index* confronts the most concentrated distribution as compared with other two measures, which are less concentrated. This indicates that when data is exposed to degree preserving perturbations, the node ranking may also change and stability problem exists to different ranges. To characterize the stability of centrality measure, we define three metrics:

**Mean Bias,** μ calculates the average of rank difference of same node from perturbed and unperturbed network for some centrality measurement and is defined as:

$$\mu = \frac{\sum_{xy} \delta_{xy}}{n(n-1)} \quad (1)$$

Where n is the size of the network and $\delta_{xy}$ is the difference between true rank of a node calculated from centrality measure on original network (x) and randomized network (y) as shown in figure 2. Lower the values of μ, stable the centrality of nodes would be.

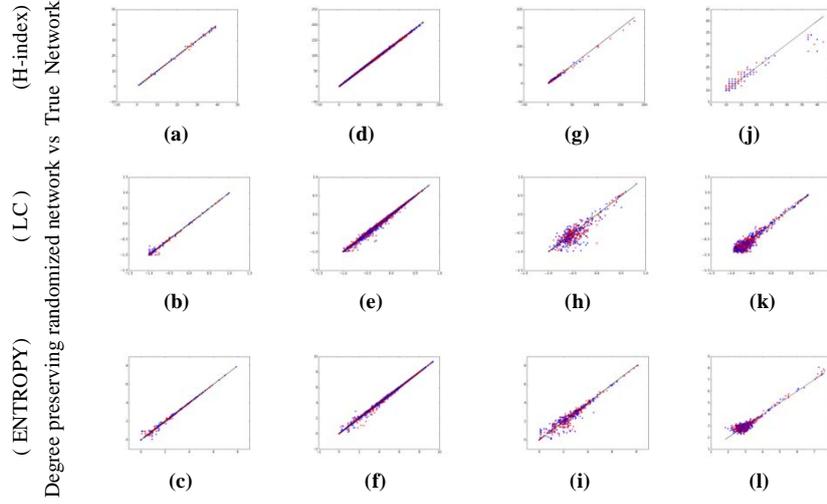

**Fig. 2** Effect of degree preserving randomizations on h-index, leverage centrality and local structure entropy. Synthetic Scale free and exponential distribution also real facebook and c elegans networks are examined. Figure shows the scatter plot of ranks in both original network and degree preserved randomized network for the same. Synthetic scale free network (a),(b),(c), Facebook (d), (e), (f), C elegans (g), (h), (i), and synthetic exponential(j), (k), (l).

**Standard Deviation of Mean Bias,** σ measures the susceptibility of ranks against the change in data occurred due to randomization. It is defined as:

$$\sigma = \sqrt{\frac{\Sigma_{xy}\,(\delta_{xy}-\mu)^2}{n(n-1)}} \quad (2)$$

High value of σ indicates that ranks of a few nodes are quite unstable and low value indicates the similar unstable level of ranks for all nodes due to randomization or unstable centrality.

**Jaccard Similarity Index** [20], measures the overlap between two rank vectors. In order to check how top ranking nodes change under different proportion of perturbation, we compare top 25 nodes according to rank given by each centrality measures for different networks. Highest value (1) indicates that the set of these nodes is not changed and lowest value (0) indicates that it has changed totally due to randomization offered to the network when compared with original vector of rank for top nodes.

## 3  Experiments and Results

For experimentation, we have two base topological configurations for network; scale free and exponential and generated synthetic data for the same also considered two real networks for each configuration as mentioned in the table II. We perform degree preserving randomization on empirical data sets. For each mentioned centrality measure we calculate the ranking vectors for perturbed and unperturbed networks. The results of mean bias μ, standard deviation σ of mean bias are shown in figure 3. It is well noticeable from the histograms for inverse mean bias and inverse standard deviation of bias score of various centrality indices, that overall H-index outperforms other centrality measurements for each topological configuration. However, all other centrality measures show up higher stability for the scale free topology of underlying network. Apparently higher values for centrality stability score of scale free configuration suggests the role of network characteristic in determining its performance under mentioned stability metrics. This fact is also supported by the work of Ghoshal and Barabasi [18], that the accomplishment achieved by page-rank algorithm in ranking the web contents is equally credited to the scale free characteristic of www. For investigating configuration of top ranked nodes in randomized networks according to different centrality metrics we use **noise model 2** as introduced by Yang [21]. Jaccard overlap of top rank nodes for all centrality measures on different datasets are evaluated with varying assortativity of the network as shown in figure 4. In majority of the cases h-index performs better than other metrics retaining top ranked node under varied level of randomization. It can be realized from the experimental evaluations that the scale free networks are more stable for local information based centrality stability performance under perturbation of network and h-index is the most stable metric when compared with present set of centrality metrics considered for evaluation.

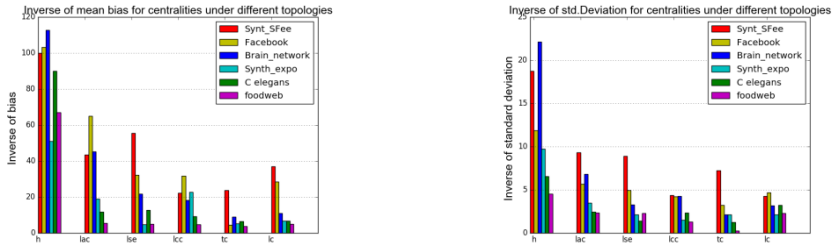

**Fig.3.** Inverse of mean bias (1/μ) and inverse of standard deviation of mean bias (1/σ) for different centrality measures under different topology networks

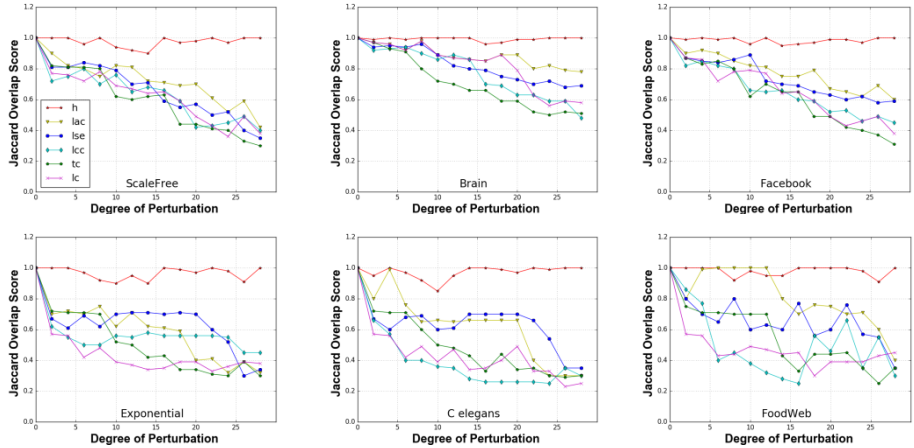

**Fig.4.** The Jaccard similarity index between the top vertices of the original and the perturbed networks for varying levels of assortativity changes with increasing number(degree) of perturbations (**noise model 2**)

## 4    Conclusion

In this work, we investigate how changing assortativity affect different centrality metrics under given topologies of the network. We find that h-index centrality outperforms other benchmark centrality metrics based on local information in the network, when network is perturbed keeping its degree distribution constant according to noise model 1. To further explore stability notion, we use Jaccard similarity index for different centrality rank vectors of top nodes when network is randomized according to noise model 2 with original top nodes vector. We find that the top rank nodes also called stable nodes are more prevalent to networks with scale free degree distribution and h-index shows high value for Jaccard overlap on majority of datasets. Our method introduces contemporary measures to analyze and evaluate stability of different centrality metrics which can be further explored to investigate other important networks parameters. This work can be enhanced by studying behavior of network with other properties like transitivity, also different noise models to effect networks could be studied in this regard.